\documentclass[prd,twocolumn,showpacs,preprintnumbers,superscriptaddress,floatfix,nofootinbib]{revtex4-1}
\usepackage{subfigure}
\usepackage{amsfonts}
\usepackage{graphicx,,booktabs}
\usepackage{amssymb}
\usepackage{amsmath,txfonts,ulem}
\usepackage{graphicx}
\usepackage{dcolumn}
\usepackage{color}
\usepackage{bm}
\usepackage{ulem}
\usepackage[colorlinks,
            citecolor=blue,
            anchorcolor=green,
            menucolor=orange,
            linkcolor=red,
            filecolor=red,
            runcolor=pink,
            urlcolor=blue,
            frenchlinks=red]{hyperref}

\allowdisplaybreaks[4]

\begin{document}

\title{Mass radius and mechanical properties of the proton via strange $\phi$ meson photoproduction}

\author{Xiao-Yun Wang}
\email{Corresponding author: xywang@lut.edu.cn}
\affiliation{Department of physics, Lanzhou University of Technology,
Lanzhou 730050, China}
\affiliation{Lanzhou Center for Theoretical Physics, Key Laboratory of Theoretical Physics of Gansu Province, Lanzhou University, Lanzhou, Gansu 730000, China}

\author{Chen Dong}
\email{dongphysics@yeah.net}
\affiliation{Department of physics, Lanzhou University of Technology,
Lanzhou 730050, China}

\author{Quanjin Wang}
\affiliation{Department of physics, Lanzhou University of Technology,
Lanzhou 730050, China}

\begin{abstract}
In this work, the cross sections of $\phi$ photoproduction are systematically investigated with the two gluon exchange model and the pomeron model. The results obtained show that the theoretical values of the two models agree well with the experimental data. Since the mass radius and mechanical properties of the proton are encoded in the scalar gravitational form factor of the momentum energy tensor, based on the differential cross sections of $\phi$ photoproduction at near-threshold predicted by the two gluon exchange model, the average mass radius of the proton is derived as
$\sqrt{\left\langle r^{2}\right\rangle _\mathrm{m} }=0.78 \pm 0.06$ fm. As a comparison, we directly extract the proton mass radius from the experimental data of $\phi$ photoproduction to obtain an average value of $0.80\pm0.05$ fm, which is very close to the result given by the two gluon exchange model. Taking a similar approach, we extracted the average value of $m_{D}$ to obtain the distribution of the pressure and shear force contributed within the proton, and compared the results with other groups. The results of this paper may provide necessary theoretical information for the subsequent in-depth study of the internal structure and properties of proton.

\end{abstract}


\maketitle

\section{Introduction}
From the quark model and QCD theory
\cite{Gross:1973id,Gross:1973ju,Gross:1974cs}, it is clear that proton can be divided into quarks and gluons. However, the mass of three of the valence quarks is only 9.6 MeV/$c^2$, which is just about one percent of the mass of a proton, and we are still exploring other sources of mass of proton. In addition, quark color confinement and asymptotic freedom are also problems that have always plagued us. These questions suggest that our understanding of the internal structure and properties of nucleons is very limited. Until recent years, it has been found that analyzing the mass radius and mechanical properties of nucleons may be an effective way to study and determine the internal structure of nucleons \cite{Kharzeev:2021qkd,Pefkou:2021fni,Wang:2021dis,Mamo:2021krl,Mamo:2022eui,Burkert:2021ith,Pefkou:2021fni}. The mass radius  is one of the essential characteristics of the proton structure, which has been proved to be strictly defined by the Gravitational Form Factors (GFF) of Energy Momentum Tensor (EMT) under the weak gravitational approximation \cite{Kharzeev:2021qkd}. In addition, EMT also contains information related to the variation of the spatial component of the space-time metric: D-term \cite{Polyakov:1999gs,Polyakov:2018zvc}. Since it can be expressed in terms of the stress tensor, which is also defined as mechanical properties \cite{Polyakov:2002yz, Burkert:2018bqq}. In Ref. \cite{Polyakov:2018zvc}, a systematic study of the internal mechanical properties of the proton through EMT was carried out. At present, the study of mass radius and mechanical properties is the focus of nuclear physics.

Since the interaction between gravitons and protons is very weak, it is difficult to directly measure the mass radius and mechanical properties of the proton through experiments. In Ref. \cite{Ji:1996ek,Ji:1996nm}, X. Ji proposed that the internal quark structure of the proton can be revealed by deeply virtual Compton scattering (DVCS). In 2018, V. D. Burkert \textit{et al}. used DVCS data to extract the pressure distribution inside the proton for the first time \cite{Burkert:2018bqq}. The strong repulsive force near the center of the proton and binding force far away from the center were perceived. Subsequently, the shear force distribution was derived with the same method \cite{Burkert:2021ith}, which discovered that most confinement forces might be around $0.6$ fm from the center of the proton.
In addition to extracting mechanical properties from DVCS data, the Lattice Quantum Chromodynamics (LQCD) and holographic QCD were developed for analyzing the mass radius and mechanical properties of the proton systematically. For example, researchers derived the GFF from LQCD and the pressure distribution inside the nucleon \cite{Shanahan:2018pib} was calculated with the light quark mass corresponding to $m_{\pi}$. In their subsequent work \cite{Pefkou:2021fni}, the mechanical properties of the proton were obtained by the scalar gravitational tripole fit, and z-expansion was involved to modify the mechanical distribution. In addition, the mass radius was calculated as $0.747\pm0.033$ fm. A novel method was developed for extracting the mass radius as $0.682$ fm by Holographic QCD in Ref. \cite{Mamo:2022eui}, and mechanical properties contained in D-term were described in detail. Besides, in 2019, the gravitational form factor of valence quarks in nucleons was calculated based on the light-cone sum rules, and the result agrees well with experimental data \cite{Anikin:2019kwi}. In 2021, the light front quark-diquark model was also proposed to deduce the scalar gravitational form factor and show the pressure and shear force distribution of the quark inside the proton \cite{Chakrabarti:2021mfd}.

Underneath a hypothesis of the scalar gravitational dipole form factor, the mass radius of the proton can be extracted with the vector quarkoniums photoproduction at the near-threshold. As well as, the scalar gravitational tripole form factor is utilized to extract the mechanical properties of the proton. In 2021, based on associating the GFF with the cross sections of $J/\psi$ photoproduction from GlueX Collaboration data \cite{GlueX:2019mkq}, the mass radius was obtained as $0.55\pm0.03$ fm \cite{Kharzeev:2021qkd} for the first time. In Ref. \cite{Wang:2021dis}, the mass radius of $0.67 \pm 0.03$ fm was extracted by utilizing the experimental cross sections of vector mesons $\phi$, $\omega$ and $J/\psi$ photoproduction. Similarly, the mass radius $0.68$ fm was derived with the identical method in Ref. \cite{Guo:2021ibg}.
However, due to the large variation of the cross section of vector mesons photoproduction at the threshold energy, and the limited accuracy and quantity of the existing experimental data, there will be some uncertainty if the mass radius is directly extracted from the photoproduction data of vector mesons. Therefore, it is necessary to introduce relevant theoretical models to calculate the cross sections of vector mesons photoproduction, and to further give the distribution of proton mass radius at the threshold energy. This allows us to compare and verify the results given by the theoretical model and the results obtained by direct extraction, so as to give a relatively reliable value of the proton mass radius. In our previous work \cite{Wang:2022vhr}, models such as two gluon exchange was used to predict the photoproduction cross sections of heavy charmonium $J/\psi$ and $\psi(2S)$. The average proton mass radius was calculated as $0.67 \pm 0.11$ fm, which is basically consistent with the values given by other groups when errors are considered. However, in further discussion, it is found that the proton mass radius value extracted from the near-threshold differential cross section of heavy quarkoniums is affected by the momentum transfer $|t|_{\min}$, so that the proton mass radius value obtained around the threshold varies greatly and unstable. Thus, the lighter vector meson $\phi$ acquired our attention.


As a traditional vector meson of $s\bar{s}$ system, $\phi$ is not only a bridge to explore the QCD from light to heavy quarks \cite{Li:2020xzs}, but also its photoproduction process is suitable for studying the mass radius and mechanical properties of the proton. As high-energy experiments progress, some experimental data of $\phi$ photoproduction near-threshold have been accumulated in recent years by CLAS, SLAC, FERMLAB, etc. \cite{Strakovsky:2020uqs,Barber:1981fj,Ballam:1972eq,Egloff:1979mg,Busenitz:1989gq,Owens:2012bv,Dey:2014tfa,LEPS:2005hax}, which provide the necessary foundation for our research. In this paper, the two gluon exchange model \cite{Ensign:1983jc,Baranov:2010zzb,Suwara:1988uz} is established based on the existing  experiment data of $\phi$ photoproduction, and the pomeron model developed by Larget \cite{Laget:1994ba} is also introduced. Since this pomeron model contains almost no free parameters and is in good agreement with the previous experiments, it can be used to compare and verify the validity of the two gluon exchange model. Based on the cross sections of $\phi$ photoproduction predicted by the both models, the mass radius $\sqrt{\left\langle r^{2}\right\rangle _\mathrm{m} }$ can be derived by GFF \cite{Guo:2021ibg,Kharzeev:2021qkd,Kou:2021qdc,Ji:1997pf,Teryaev:2016edw}, and proton mechanical properties including pressure and shear force distributions will be analyzed. Moreover, the results of this paper can provide theoretical reference for further study of the proton structure with the Electron-Ion Collider (EIC) facilities
\cite{Anderle:2021wcy,Accardi:2012qut}.

The outline of this work is organized as follows. The formalism of the two gluon exchange and the pomeron models are introduced, and the mass radius and mechanical properties of the proton are described in Sec. \ref{sec2}. The results of fitting and proton coherence are presented in Sec. \ref{sec3}. A general summary in Sec. \ref{sec4}.

\section{Formalism}\label{sec2}
\subsection{Two gluon exchange model}

\begin{figure}[htbp]
	\begin{center}
		\includegraphics[scale=0.4]{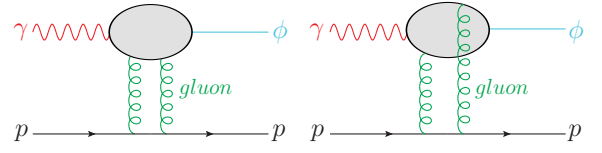}
		\caption{ The Feynman diagram for $\phi$ production based on the two gluon exchange model.}  \label{fig:1}
	\end{center}
\end{figure}

In the two gluon exchange model, the photon fluctuation into the quark-antiquark pair ($\gamma \rightarrow q + \bar{q}$) is described explicitly in Fig. \ref{fig:1}. To begin with, the photon splits into a dipole ($s\bar{s}$). Furthermore, the dipole exchange two gluons to scatter off the initial proton. Eventually, the dipole formation of the final state $\phi$, as well as the initial proton, preserves the proton interaction as two gluons are colorless. In brief, the two gluon exchange process is dominated by $\phi$ production near the threshold.

This work considers the $\phi p$ sattering process $\gamma p\rightarrow\phi p$ with two gluon exchange model. In lowest order perturbative QCD, the amplitude of $\phi$ photoproduction can be derived from Ref. \cite{Brodsky:1994kf,Ryskin:1992ui,Ryskin:1995hz},
\begin{equation}
	\begin{aligned}
	{\cal T}=&\frac{i 2 \sqrt{2} \pi^{2}}{3} m_{s} \alpha_{s} e_{s} f_{V} F_{2 g}(t) \\
		&\int d l^{2} P_{g}^{2}(l)\left[P_{+}(l)-P_{-}(l)\right] G(l).
	\end{aligned}  \label{eq:1}
\end{equation}
where $m_{s} = 0.486$ GeV and $e_{s} = -1/3$ are the mass \cite{Kou:2021bez} and charge of quark, respectively.
$f_{v}$ is the pion decay constant, and it can be extracted from the leptonic decay width, which is given by,
\begin{equation}
	\Gamma_{e^{+} e^{-}}=\frac{4 \pi \alpha^{2}}{3 m_{\phi}} f_{V}^{2}.
\end{equation}
The $\Gamma_{e^{+} e^{-}} = 1.27$ keV is the decay width of $\phi$ which can be obtained from Ref. \cite{Sibirtsev:2004ca}, and $\alpha = 1/137$. With this we can calculate $f_{V} =0.0761$ GeV.
In Eq.(1), $P_{g}(l)$ is the gluon propagator and it can be taken as $1/l^{2}$. $P_{+}(l)$ is the propagator of the off-shell quark that two gluons coupled to the same quarks of the vector meson is written as,
\begin{equation}
	P_{+}(l)=\left(-2 m_{s}^{2}\right)^{-1},
\end{equation}
and $P_{-}(l)$ is the propagator of the off-shell quark that two gluons coupled to different quarks of the vector meson is given by,
\begin{equation}
	P_{-}(l)=\left(-2 m_{s}^{2}-2 l^{2}\right)^{-1}.
\end{equation}
The factor $F_{2 g}(t)$ in Eq. (\ref{eq:1}) is interpreted as the dependence of the amplitude dependence of the two gluons in the proton is expressed as \cite{Ryskin:1992ui,Xu:2020uaa},
\begin{equation}
	F_{2 g}(t)=\frac{4 m_{p}^{2}-2.8 t}{4 m_{p}^{2}-t} \frac{1}{\left(1-t / t_{0}\right)^{2}}.
\end{equation}
where $t_{0} = 0.71$ GeV$^{2}$ and $m_{p}$ is the proton mass. The probability that a dipole captures a gluon with momentum $l$ from a proton is defined by $G(l)$. The integral of $G(l)$ is affiliated with the gluon distribution function $xg(x,Q^{2})$ \cite{Ryskin:1992ui,Brodsky:1994kf},
\begin{equation}
	xg\left(x, Q^{2}\right)=\int d l^{2} \frac{G(l)}{l^{2}}.
\end{equation}
From the above discussion, the amplitude of the two gluon exchange can be constructed as,
\begin{equation}
	\begin{aligned}
		{\cal T}=& \frac{i \sqrt{2} \pi^{2}}{3} m_{s} \alpha_{s} e_{s} f_{V} F_{2 g}(t)\left[\frac{xg\left(x, Q_{0}^{2}\right)}{m_{s}^{4}}\right.\\
		&\left.+\int_{Q_{0}^{2}}^{+\infty} \frac{d l^{2}}{m_{s}^{2}\left(m_{s}^{2}+l^{2}\right)} \frac{\partial  xg\left(x, l^{2}\right)}{\partial l^{2}}\right]   \label{eq:7}.
	\end{aligned}
\end{equation}
$d\sigma / dt = \alpha|{\cal T}|^{2}$ is the differential cross section with the normalized amplitude. Then, the differential cross section, in the lowest order is denoted as \cite{Xu:2020uaa},
\begin{equation}
	\label{eq:8}
	\frac{d \sigma}{d t}=\frac{\pi^{3} \Gamma_{e^{+} e^{-}} \alpha_{s}}{6 \alpha m_{s}^{5}}\left[xg\left(x, m_{\phi}^{2}\right)\right]^{2} \exp (b_{0} t),
\end{equation}
where $x = m^{2}_{\phi}/W^{2}$, $\alpha_{s} = 0.701$ is the QCD coupling constant from Ref. \cite{Kou:2021bez}. The $t$-dependence of $\phi$ photoproduction can be exponentially parameterized and $b_{0}$ is constant slope. In addition, the $|t|$-dependence of exponential slope $b_{0}$ also can be calculated via $b_{0} = \dfrac{d}{dt}\ln[\dfrac{d\sigma}{dt}]$.

One note that the differential and total cross sections formula obtained primarily depends on the gluon distribution function $xg(x,m^{2}_{\phi})$. For example, in Ref. \cite{Goloskokov:2007zb}, the gluon distribution function given by CTEQ6M \cite{Pumplin:2002vw} can achieve a good description of the cross section of light vector mesons electroproduction. However, for the $\phi$ meson photoproduction near threshold, it is difficult to interpret the experimental data well using the gluon distribution functions given in the current literature \cite{Ball:2011uy,Owens:2012bv,Accardi:2016qay,Wang:2016sfq}.
In this work, we simplified parameterized gluon distribute function $xg(x,m^{2}_{\phi}) = A_{0}x^{A_{1}}(1-x)^{A_{2}}$ with three parameters $A_{0},A_{1},A_{2}$ which can be obtained by the experimental data of $\phi$ photoproduction.

To get the total cross section, one needs to integrate the Eq. (\ref{eq:8}) from $t_{min}(W)$ to $t_{max}(W)$ over the allowed kinematical range,
\begin{equation}
	t_{min}(W)(t_{max}(W)) = [\dfrac{m_{1}^{2} - m_{3}^{2} - m_{2}^{2} + m_{4}^{2}}{2W}]^{2} - (P_{1cm} \mp P_{3cm})^{2},
\end{equation}
with
\begin{equation}
	P_{icm} = \sqrt{E_{icm}^{2} - m_{i}^{2}}(i = 1, 3),
\end{equation}
\begin{equation}
	E_{1cm} = \dfrac{W^{2} + m_{1}^{2} - m_{2}^{2}}{2W}, E_{3cm} = \dfrac{W^{2} + m_{3}^{2} - m_{4}^{2}}{2W}.
\end{equation}
Then, the total cross section of $\phi$ photoproduction as,
\begin{equation}
	\sigma= \frac{\pi^{3} \Gamma_{e^{+} e^{-} \alpha_{s}}}{6 \alpha m_{s}^{5}}\left[x g\left(x, m_{\phi}^{2}\right)\right]^{2} \int_{t_{min}(W)}^{t_{max}(W)} e^{b(W) t} d|t|.
\end{equation}
The exponential slope $b(W)$ is proportional to the centre of mass energy $W$. A study by S. Chekanov shows that in the high-energy region the form of $b(W)$ can be derived as \cite{Close:2002ky},
\begin{equation}
	b(W)=b_{0}+0.46 \cdot \operatorname{In}\left(W / W_{0}\right).
\end{equation}
where $W_{0} = 2.29$ GeV is a modified value to calculate $b(W)$ at near-threshold energy, and $b_{0}$ will be obtained by fitting experimental data of $\phi$ photoproduction.
\subsection{Pomeron model}
\begin{figure}[htbp]
	\begin{center}
		\includegraphics[scale=0.4]{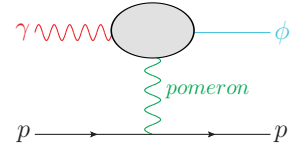}
		\caption{ The Feynman diagram for $\phi$ production based on the pomeron model.}  \label{fig:2}
	\end{center}
\end{figure}

The pomeron model \cite{Laget:1994ba} also describes the photoproduction cross section of vector mesons in attribute, scattering $q\bar{q}$ state by protons and finally transforming the $q\bar{q}$ state into vector mesons. The interaction between photon and proton in the pomeron model is elaborated in Fig. \ref{fig:2}, and the differential cross section of meson $\phi$ can be written as,
\begin{equation}
	\frac{\mathrm{d} \sigma}{\mathrm{d} t}=\frac{81 m_{\mathrm{\phi}}^{3} \beta^{4} \mu_{0}^{4} \Gamma_{\mathrm{e}^{+} \mathrm{e}^{-}}}{\pi \alpha}\left(\frac{s}{s_{0}}\right)^{2 \alpha(t)-2}\left(\frac{F(t)}{\left(m_{\mathrm{\phi}}^{2}-t\right)\left(2 \mu_{0}^{2}+m_{\mathrm{\phi}}^{2}-t\right)}\right)^{2},
\end{equation}
with the form factor $F(t)$ is,
\begin{equation}
	F(t) = \frac{4m_{p}^{2}-2.8t}{(4m_{p}^{2}-t)(1-\frac{t}{0.7})^{2}}.
\end{equation}
and the Regge trajectory $\alpha(t)=1.08+0.25t$ is derived from Ref. \cite{Landshoff:1990kj}, where coefficient $\beta=2$ GeV$^{-1}$, $\mu_{0}^{2}=1.1$ GeV$^{2}$ and $s_{0}=4$ GeV$^{2}$.
\subsection{Mass radius of proton}
The mass radius of the proton describes the mass distribution within the proton, which can be summarized as the mass density distribution. Generally, with the weak field approximation, the mass distribution, D-term, and spin we are interested can be obtained by GFF \cite{Ji:1996nm,Ji:1997pf,Kharzeev:2021qkd,Wang:2021dis,Guo:2021ibg,Teryaev:2016edw}. The mass radius is defined as the derivative of momentum transfer to GFF, which can be written as \cite{Kharzeev:2021qkd},
\begin{equation}
\left\langle R_{\mathrm{m}}^2\right\rangle=\left.\frac{6}{M} \frac{d G(t)}{d t}\right|_{t=0},\label{18}
\end{equation}
where $G(t)=M/(1-t/m_{r}^{2})^{2}$ is the dipole form of scalar GFF and is normalized. Therefore, we can summarize the form of mass radius as,
\begin{equation}
\left\langle R_{\mathrm{m}}^2\right\rangle=\frac{12}{m_r^2}.\label{19}
\end{equation}
where $m_{r}$ is a free parameter can be obtained by fitting the quarkonium photoproduction data. The photoproduction of quarkonium near the threshold is susceptible to scalar GFF and mass distribution. In the small momentum range, the photoproduction differential cross section of quarkonium can be described by the scalar GFF, which can be written as $d \sigma / d t \propto G^{2}(t)$ \cite{Kharzeev:2021qkd,Wang:2021dis}. Consequently, it is available to extract the mass radius of the proton from the cross section of quarkonium photoproduction.
\subsection{Mechanical properties of proton}
The mechanical properties of the proton, including the pressure and shear force distribution, are also the direction of proton internal structure research. These properties are encoded in Quantum Chromodynamics Energy Momentum Tensor (EMT). The EMT satisfies the conservation law and can be described as the space-time symmetric translation related conservation current based on Noether’s Theorem, which can be represented as \cite{Kou:2021qdc},
\begin{equation}
	\begin{aligned}
		&\partial^{\mu} \hat{T}_{\mu \nu}=0 \\
		&\hat{T}_{\mu \nu}=\sum_{q} \hat{T}_{\mu \nu}^{q}+\hat{T}_{\mu \nu}^{g}
	\end{aligned}
\end{equation}
where $\hat{T}_{\mu \nu}^{q}$ and $\hat{T}_{\mu \nu}^{g}$ represent the contribution of quarks and gluons, respectively. The main object of this work is the proton related problem. For the proton with spin-$\frac{1}{2}$, the one particle state can be represented by $|\boldsymbol{p}\rangle$, and the normalization condition is $\left\langle p^{\prime} \mid p\right\rangle=2 p^{0}(2 \pi)^{3} \delta^{(3)}\left(\boldsymbol{p}^{\prime}-\boldsymbol{p}\right)$. So, the matrix element of EMT for proton can be introduced as \cite{Polyakov:2018zvc},
\begin{equation}
	\begin{aligned}
		&\left\langle p^{\prime}, s^{\prime}\left|\hat{T}_{\mu \nu}^{a}(x)\right| p, s\right\rangle=\bar{u}^{\prime}\left[A^{a}(t) \frac{\gamma_{\{\mu} P_{\nu\}}}{2}\right. \\
		&+B^{a}(t) \frac{i P_{\{\mu} \sigma_{\nu\} \rho} \Delta^{\rho}}{4 m}+D^{a}(t) \frac{\Delta_{\mu} \Delta_{\nu}-g_{\mu \nu} \Delta^{2}}{4 m} \\
		&\left.+m \bar{c}^{a}(t) g_{\mu \nu}\right] u e^{i\left(p^{\prime}-p\right) x},
	\end{aligned} \label{eq:19}
\end{equation}
with the notation,
\begin{equation}
a_{\{\mu} b_{\nu\}}=a_{\mu} b_{\nu}+a_{\nu} b_{\mu},
\end{equation}
Where $\bar{u}(p, s) u(p, s)=2 m$ is spin normalization, $A^{a}(t)$, $B^{a}(t)$, $D^{a}(t)$, and $\bar{c}^{a}(t)$ represent the form factors of individual quarks and gluons, and these four form factors depend on renormalization scale. The $B^{a}(t)=0$ for gluons by calculation of lattice QCD \cite{Alexandrou:2017oeh}.
The  tripole parametrization for D-term from perturbation theory is,
\begin{equation}
D(t)=\frac{D(0)}{\left(1-t / m_{D}^{2}\right)^{3}}. \label{eq:21}
\end{equation}
where $m_{D}$ and $D(0)$ are free parameters.

The $D$-term contains information about the distribution of internal pressure $p(r)$ and shear force $s(r)$ of the proton. Using the Eq. (\ref{eq:21}) and making $t=-q^{2}$, then the tripole parameterization $D(t)$ can be rewritten as $D(-q^{2})=D(0)/(1+q^{2}/m_{D}^2)^{3}$. The D-term of the three-dimensional coordinates can be obtained from the Fourier transform of $D(q)$ \cite{Polyakov:2018zvc} with $E=m_{N}$,
\begin{equation}
\begin{aligned}
\tilde{D}(r)&=D(0) \int \frac{d^{3} q}{2 m_{N}(2 \pi)^{3}} \frac{e^{-i q \cdot r}}{\left(1+\frac{q^{2}}{m_{D}^{2}}\right)^{3}}\\
&=\int_{0}^{\pi} \int_{0}^{2 \pi} \int_{0}^{+\infty} \frac{q^{2} \sin (\theta)}{(2 \pi)^{3}2m_{N}} \frac{e^{-i q r \cos (\theta)} D(0)}{\left(1+q^{2} / m_{D}^{2}\right)^{3}} d \theta d \phi d q \\
&=\frac{D(0)}{4 \pi^{2} r m_{N}} \int_{0}^{+\infty} q \frac{\sin (q r)}{\left(1+q^{2} / m_{D}^{2}\right)^{3}} d q \\
&=\frac{D(0)}{64 \pi m_{N}} e^{-m_{D} r} m_{D}^{4}\left(\frac{1}{m_{D}}+r\right)\\
\end{aligned} \label{eq:22}
\end{equation}
Therefore, $p(r)$ and $s(r)$ can be calculated by Eq. (\ref{eq:22})
\begin{equation}
\begin{aligned}
p(r)&=\frac{1}{3} \frac{1}{r^{2}} \frac{d}{d r} r^{2} \frac{d}{d r} \bar{D}(r)\\
&=\frac{D(0)}{192 \pi m_{N}}\left(m_{D} r-3\right) m_{D}^{5} e^{-m_{D}r}
\end{aligned}\label{eq:23}
\end{equation}

\begin{equation}
\begin{aligned}
s(r)&=-\frac{1}{2} \frac{d}{d r} \frac{1}{r} \frac{d}{d r} \bar{D}(r)\\
&=-\frac{D(0)}{128 \pi m_{N}} m_{D}^{5} r e^{-m_{D}r}
\end{aligned}\label{eq:24}
\end{equation}

\section{RESULTS}\label{sec3}

The parameterized gluon distribution function  $xg(x, m_{\phi}^{2}) = A_{0} x^{A_{1}} (1-x)^{A_{2}}$ has been introduce in the previous section. $A_{0}, A_{1}, A_{2}$ and $b_{0}$ are derived by combining a global analysis of the total cross section of $\phi$ photoproduction from Ref. \cite{Strakovsky:2020uqs,Barber:1981fj,Ballam:1972eq,Egloff:1979mg,Busenitz:1989gq,Owens:2012bv} and the differential cross section ($W =  1.98 \sim 2.30$ GeV) by CLAS \cite{Dey:2014tfa} and LEPS \cite{LEPS:2005hax} collaboration which are listed in Tab. \ref{tab:table1}. The fitted $b_{0}=3.60\pm0.04$ GeV$^{-2}$ is near to $3.4$ GeV$^{-2}$ obtained by exponential slope $I \sim e^{-b_{0}(t-t_{min})}$ based on the pomeron exchange model mentioned in the Ref. \cite{CLAS:2013jlg,Titov:2003bk}, which adequately demonstrates the effectiveness of the two gluon exchange model.
The total and differential cross sections of the two gluon exchange model can be rewritten with the gluon distribution function $xg(x, m_{\phi}^{2}) = 0.36 x^{-0.055} (1-x)^{0.12}$.
\begin{table} \small
	\caption{\label{tab:table1}
	The slope $b_{0}$ and parameters $A_{0}, A_{1}, A_{2}$ of the gluon distribution function are obtained by fitting the experimental data and $\chi^{2}/$d.o.f value.}
	\begin{ruledtabular}
		\begin{tabular}{ccccc}
			$A_{0}$&$A_{1}$&$A_{2}$&$b_{0}$ (GeV$^{-2}$) &$\chi^{2}/$d.o.f\\
			\hline
			$0.36 \pm 0.04$&$-0.055 \pm 0.003$ & $0.12 \pm 0.03$ &$3.60\pm0.04$ &$2.87$\\
		\end{tabular}
	\end{ruledtabular}
\end{table}
\begin{figure}[htbp]
	\begin{center}
		\includegraphics[scale=0.4]{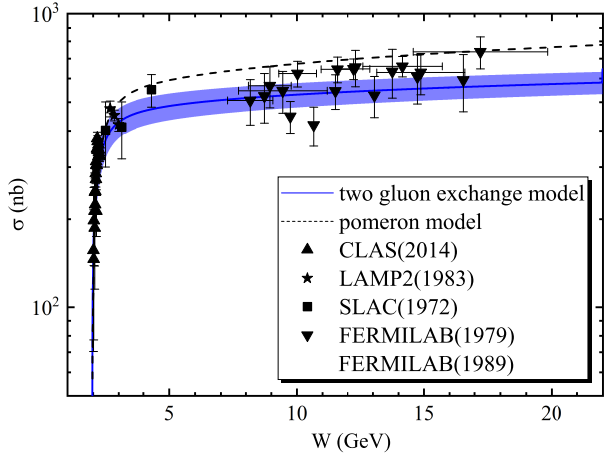}
		\caption{ The total cross section of the channel $\gamma p \rightarrow \phi p$ as a function of center of mass energy $W$. The line (blue) and dashed (black) lines are for the two gluon exchange model and pomeron model, respectively. Here, taking the error of $A_{0}$ as the error band.}\label{fig:3}
	\end{center}
\end{figure}

\begin{figure}[htbp]
	\begin{center}
		\includegraphics[scale=0.43]{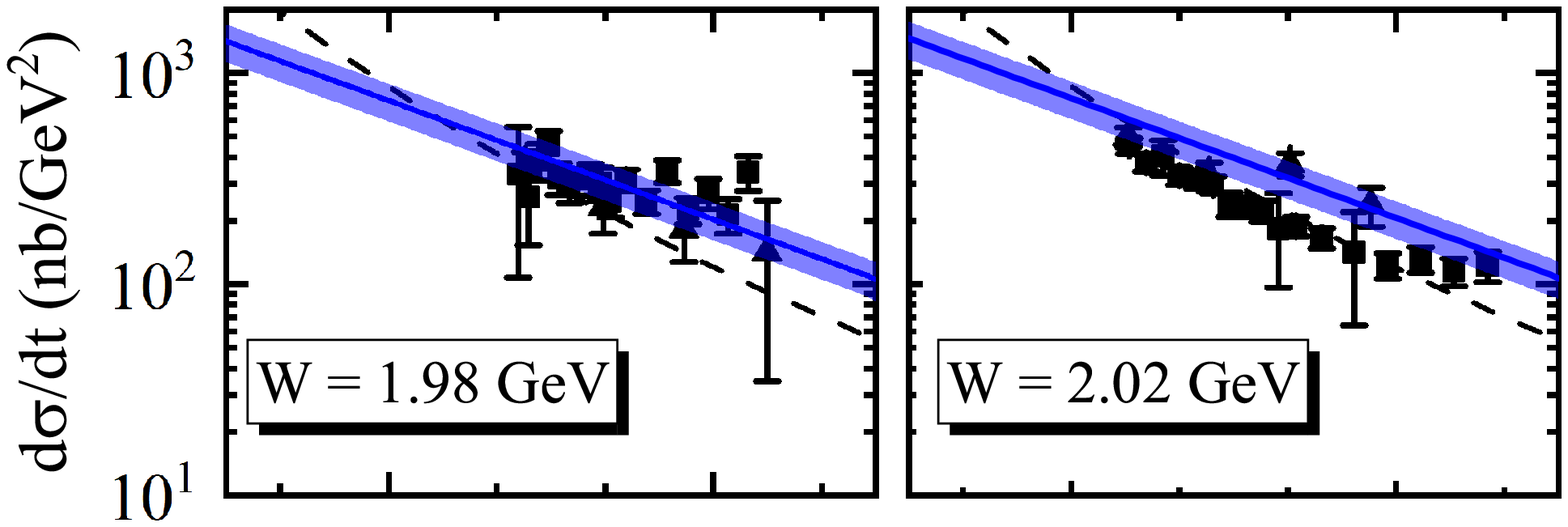}
		\\
		\includegraphics[scale=0.43]{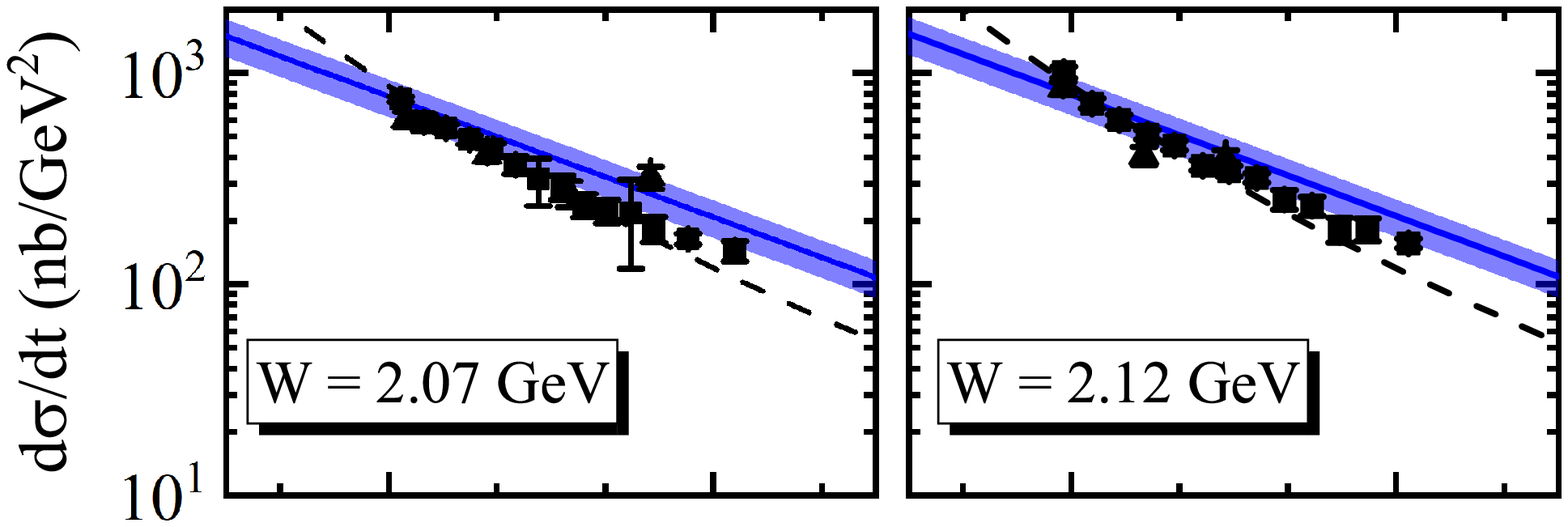}
		\\
		\includegraphics[scale=0.43]{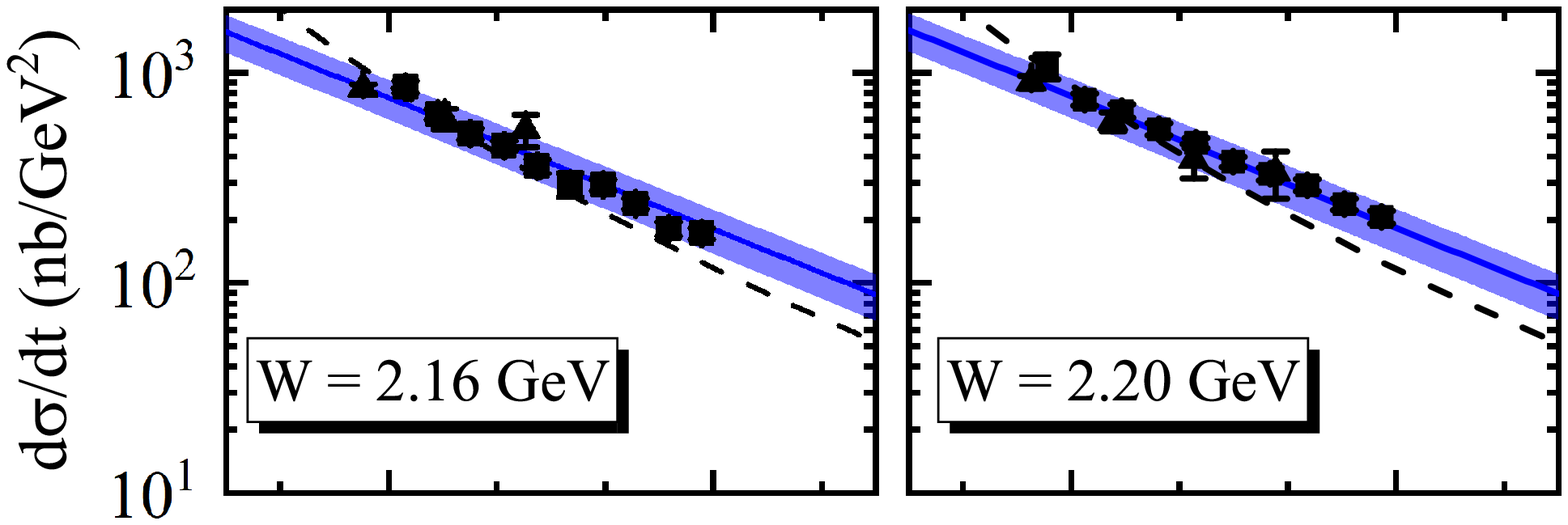}
		\\
		\includegraphics[scale=0.43]{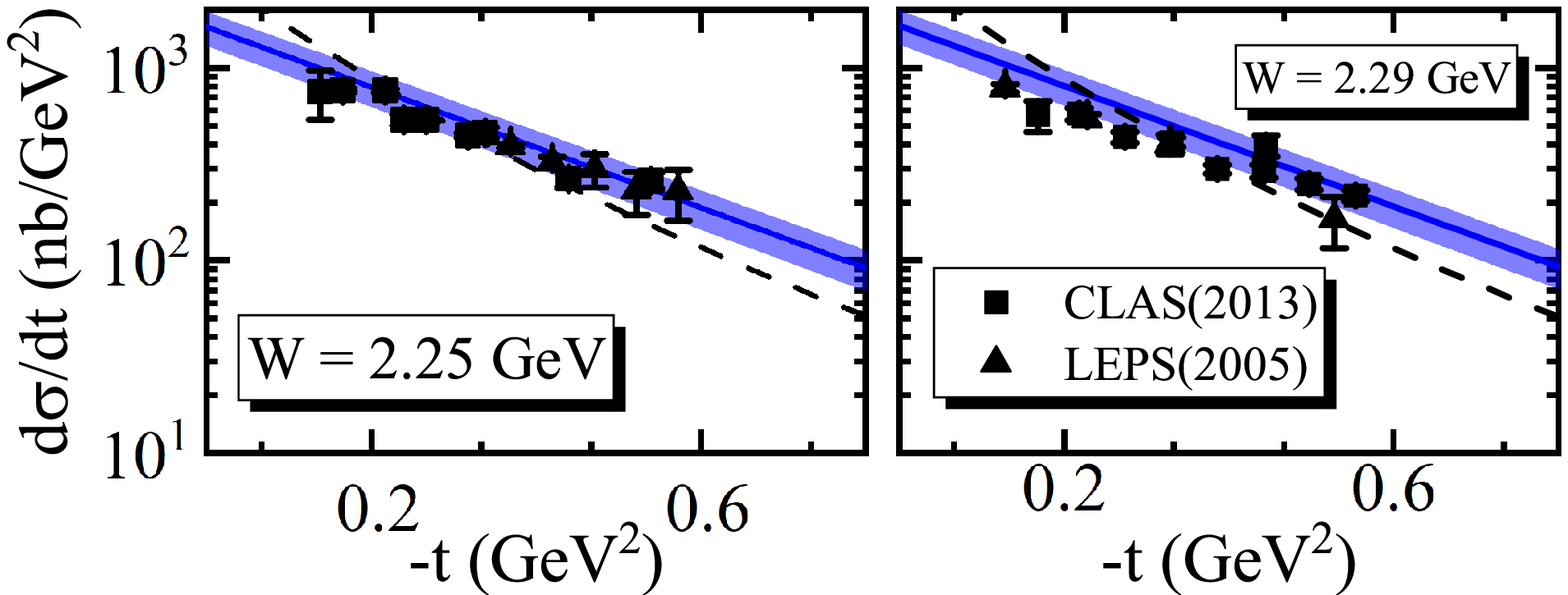}
		
		\caption{ The differential cross sections of the channel $\gamma p \rightarrow \phi p$ as a function of $-t$ at different $W$ values. Here, The blue solid line and the black dashed line correspond to the two gluon exchange model and the pomeron model, respectively.}\label{fig:4}
	\end{center}	
\end{figure}

The cross sections of $\phi$ photoproduction at the near-threshold are predicted as shown in Fig. \ref{fig:3} and Fig. \ref{fig:4}. One can figure the two gluon exchange model shows satisfactory agreement with experimental data. And $\chi^{2}/$d.o.f is calculated to be $2.87$, indicating that the fit is in line with expectations. But $\chi^{2}/$d.o.f is not tiny enough, mainly because the error bar measured in the experiment is large. To demonstrate the availability of $\phi$ photoproduction based on the two gluon exchange model, the pomeron model without any fitting parameters is introduced which was developed by Larget in Ref. \cite{Laget:1994ba}. One can conclude that the two gluon exchange model and the pomeron model almost overlap at the location nearest to the threshold in Fig. \ref{fig:3}. However, the differential cross section of $\phi$ fitted by the two gluon exchange model is more efficient than the pomeron model as shown in Fig. \ref{fig:4}.
Combined with our previous work \cite{Wang:2022vhr,Zeng:2020coc}, the total cross section of $J/\psi$ and $\psi(2S)$ were predicted based on the two gluon exchange model, which is in agreement with the experimental data. Therefore, the current results can sufficiently illustrate the validity of the cross sections of vector mesons photoproduction at the near-threshold predicted via the two gluon exchange model.

\begin{figure}[htbp]
	\begin{center}
		\includegraphics[scale=0.4]{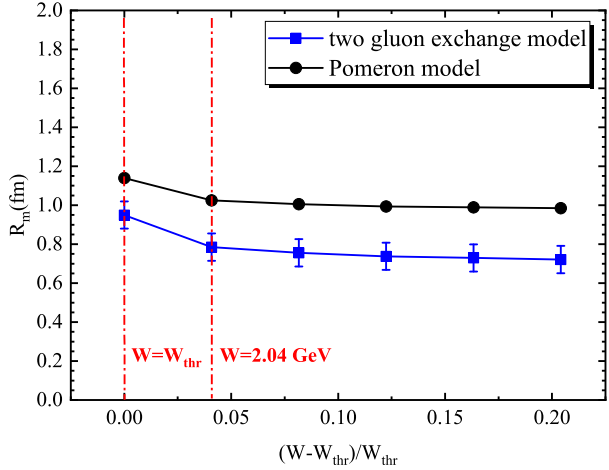}
		\caption{ The mass radius of the proton from differential cross sections of $\phi$ photoproduction at  the center of mass energy $W \in[1.96,2.36]$ GeV, $\Delta W = 0.08$ GeV. The average is $0.78 \pm 0.06$ fm and $1.02 \pm 0.01$ fm, corresponding to the two gluon exchange and pomeron models, respectively.}   \label{fig:5}
	\end{center}	
\end{figure}

\begin{figure}[htbp]
	\begin{center}
		\includegraphics[scale=0.4]{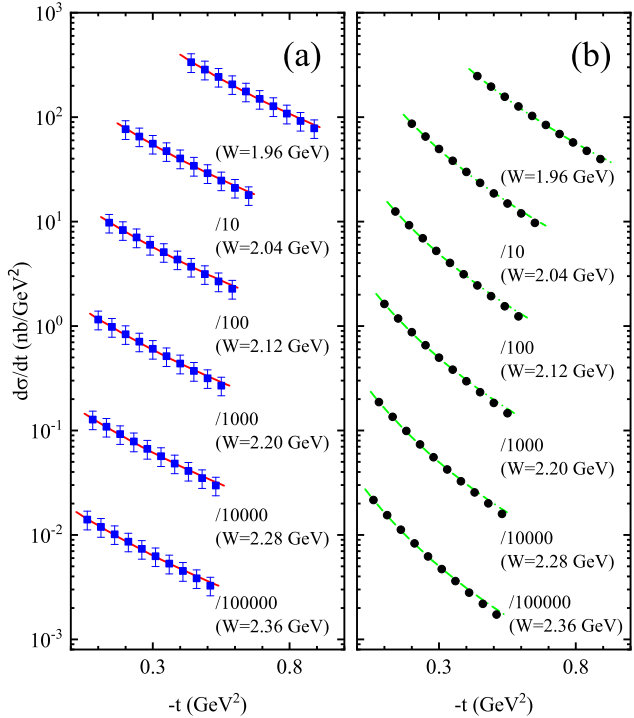}
		\caption{The fitting of $d \sigma/d t$ and $G(t)$ based on the two gluon exchange model (a) and pomeron model (b).}   \label{fig:6}
	\end{center}	
\end{figure}

Ensure that the consequences are persuasive, the mass radius is obtained via the differential cross sections of the $\phi$ photoproduction predicted at $W \in [1.96,2.36]$ with $\Delta W=0.08$ GeV, and the range of $|t|$ from $t=|t_{min}|$ select ten points with $\Delta |t|=0.05$ GeV$^{2}$. Then, the average mass radius is derived as $0.78 \pm 0.06$ fm and $1.02 \pm 0.01$ fm, corresponding to the two gluon exchange model and the pomeron model, respectively. The trend of mass radius with center of mass energy based on the two models as shown in Fig. \ref{fig:5} and
Fig. \ref{fig:6} shows the fitting of $d \sigma/d t$ and $G(t)$.
A noticeable decline for both models around the threshold $W=1.96$ GeV, while the trend gradually decreases with the increase of energy. After $W=2.04$ GeV, the trend tends to be flat. The condition is primarily due to the influence of momentum transfer $|t_{min}|$, and a similar decline also appears in the heavy charmoniums $J/\psi$ and $\psi(2S)$. However, the light meson $\phi$ is less affected by $|t_{min}|$, so the trend decreases smaller than the heavy vector meson at the threshold.


\begin{figure}[htbp]
	\begin{center}
		\includegraphics[scale=0.4]{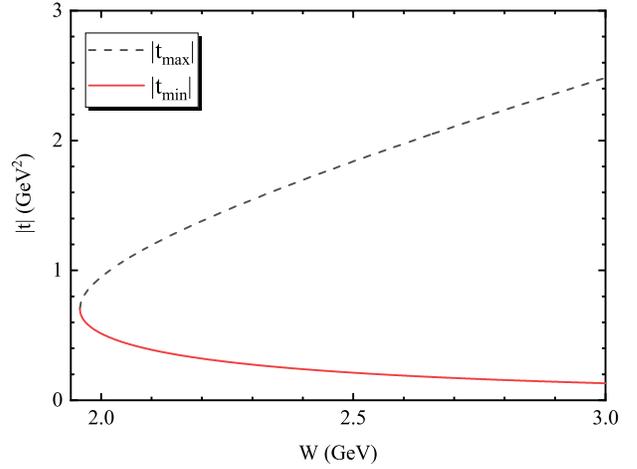}
		\caption{The change of $|t|$ with center of mass energy $W$ is near the photoproduction threshold of $\phi$.}   \label{fig:7}
	\end{center}	
\end{figure}

Although $|t_{min}|$ has less influence on light vector mesons than on heavy vector mesons, there still exists some variation with the increase of center of mass energy $W$, as shown in Fig. \ref{fig:7}.
With an increase of energy, the impact of $|t_{min}|$ on the differential cross section decreases. Away from the threshold, it can be approximated as $|t_{min}| \rightarrow 0$. If the points at the threshold are not considered, the average mass radius is calculated at $W \in[2.04,2.36]$ GeV as $\sqrt{\left\langle r^{2}\right\rangle _\mathrm{m} }=0.75$ and $\sqrt{\left\langle r^{2}\right\rangle _\mathrm{m} }=1.00$ fm, corresponding to the two gluon exchange model and the pomeron model, respectively.
By comparing the results obtained above, the average radius of ignoring the threshold is only $0.02\sim0.03$ fm smaller, which means that $d \sigma_{\gamma p \rightarrow \phi p} / d t$ at the threshold  has little effect on the average radius. This situation is in intense contrast to the heavy charmoniums $J/\psi$ and $\psi(2S)$ \cite{Wang:2022vhr}, which proves that the mass radius from the cross sections of $\phi$ photoproduction at the near-threshold is more stable.

\begin{figure}[htbp]
	\begin{center}
		\includegraphics[scale=0.43]{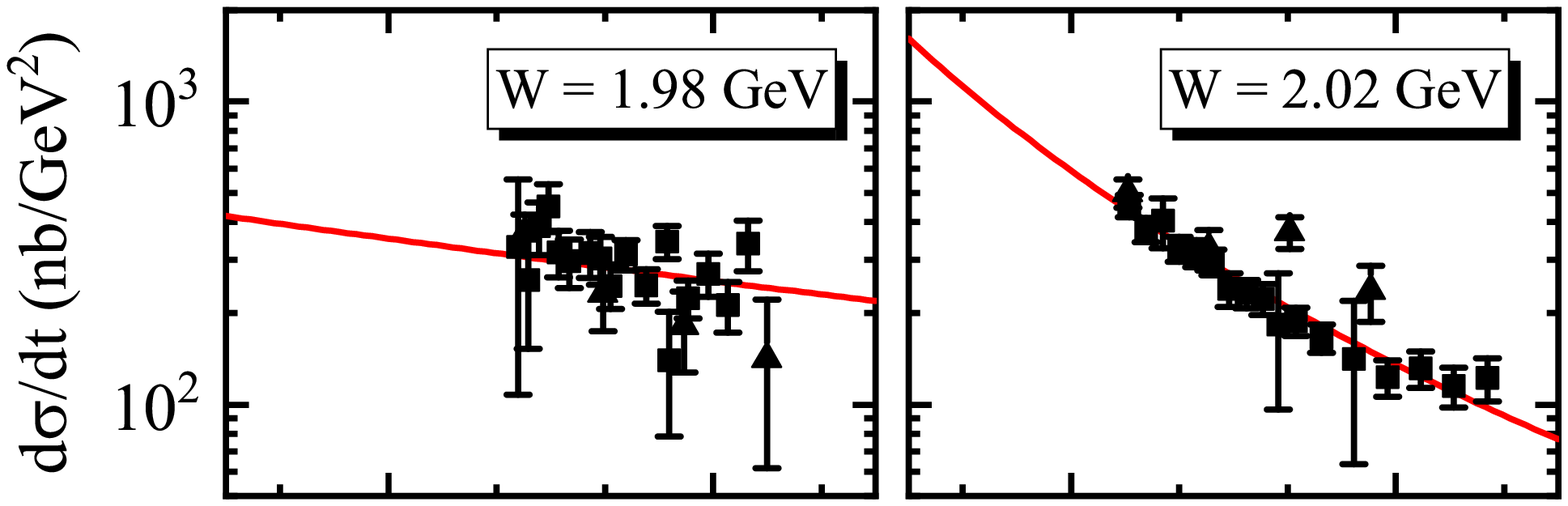}
		\\
		\includegraphics[scale=0.43]{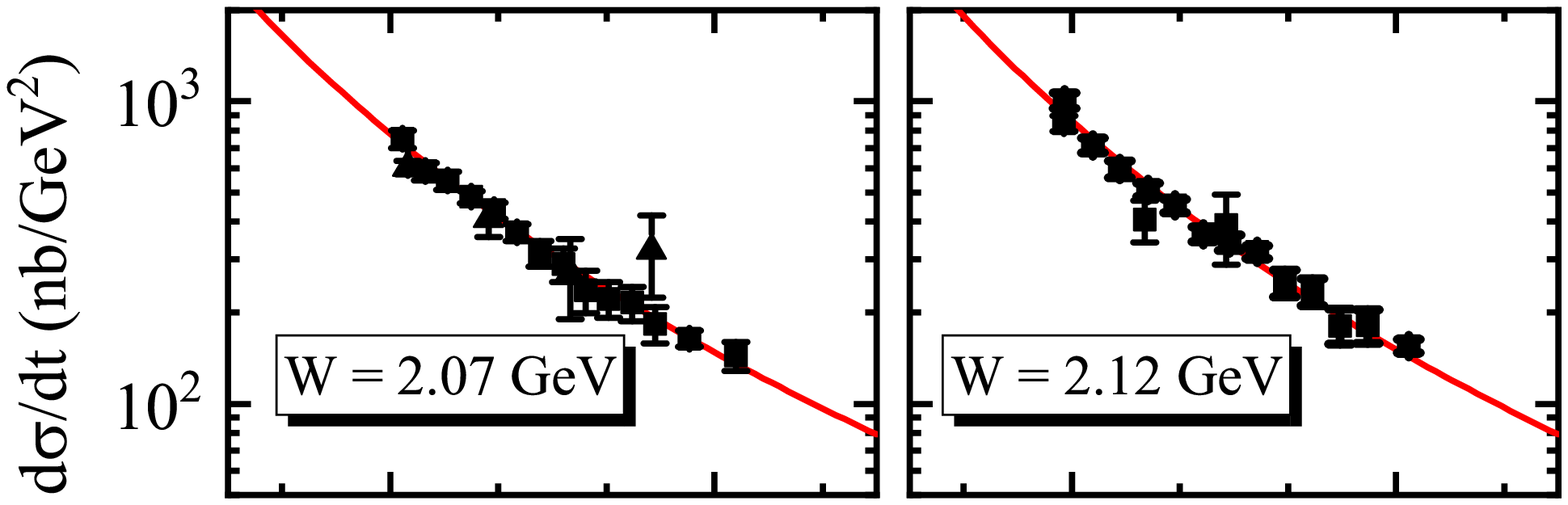}
		\\
		\includegraphics[scale=0.43]{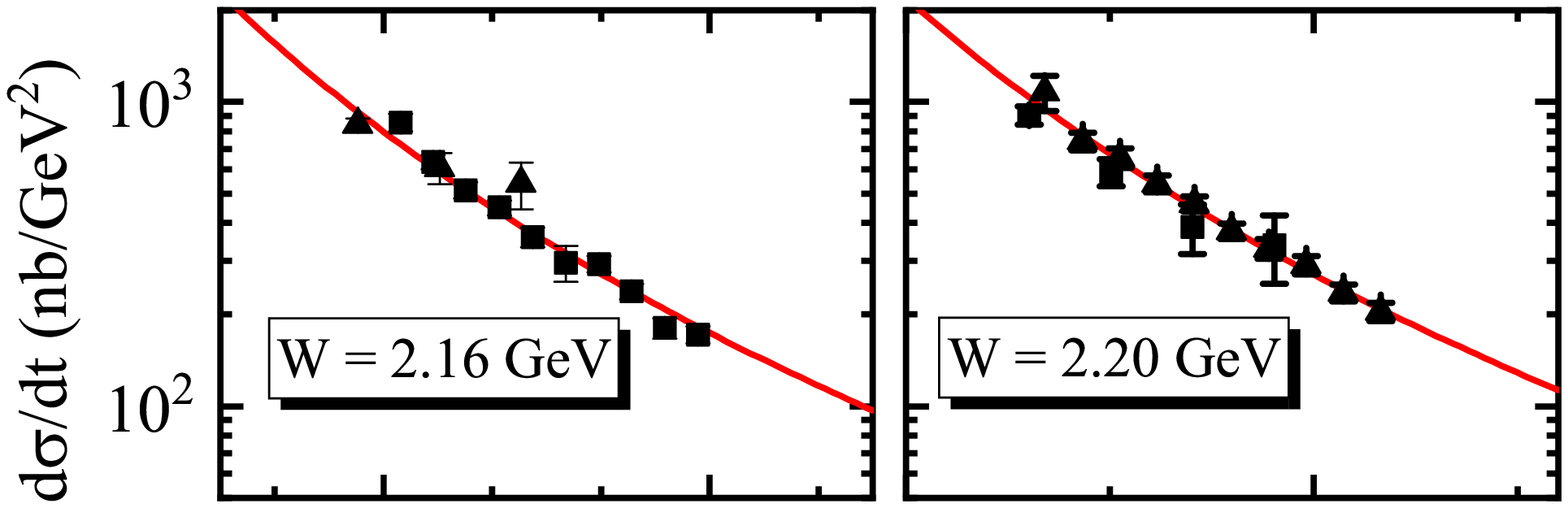}
		\\
		\includegraphics[scale=0.43]{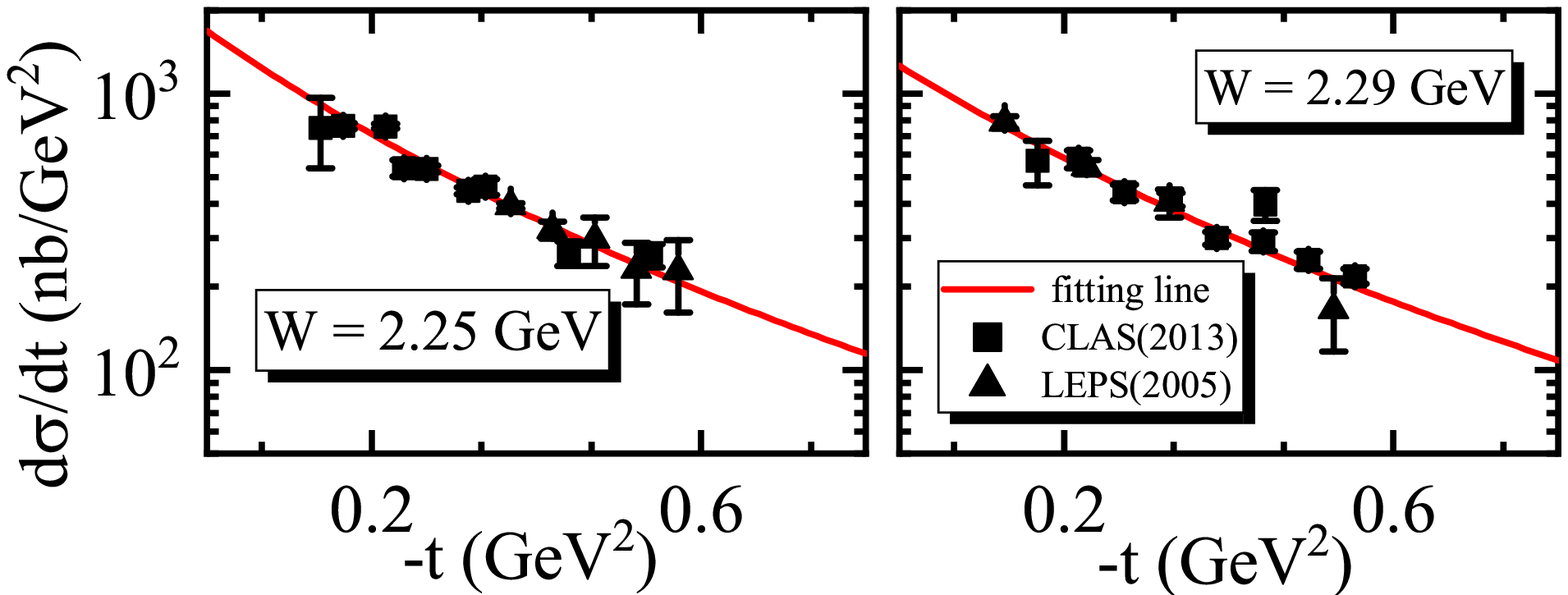}
		\caption{ The fitting of $d \sigma/d t$ and $G(t)$ based Based on the differentiation experiment cross section from CLAS \cite{Dey:2014tfa} and LEPS \cite{LEPS:2005hax} collaboration.}\label{fig:9}
	\end{center}	
\end{figure}

\begin{figure}[htbp]
	\begin{center}
		\includegraphics[scale=0.4]{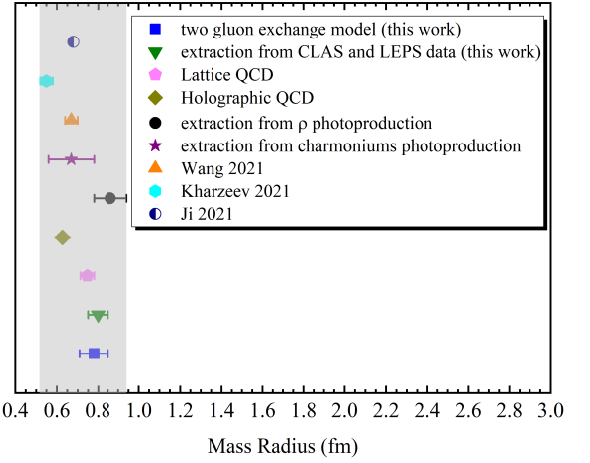}
		\caption{ Comparison of the results of proton mass radius $\sqrt{\left\langle r^{2}\right\rangle _\mathrm{m} }$ with several groups. The average based on the two gluon exchange model is $0.78 \pm 0.06$ fm. The average value of proton mass radius extracted directly from the near-threshold cross section of $\phi$ photoproduction is $0.80\pm0.05$ fm.
		The result of $0.55 \pm 0.03$ fm is derived from Ref. \cite{Kharzeev:2021qkd} and
		$0.67 \pm 0.03$ fm is the results of Ref. \cite{Wang:2021dis}. The results of $0.747 \pm 0.033$ fm and $0.682 \pm 0.012$ fm are developed by Lattice QCD \cite{Pefkou:2021fni} and Holographic QCD \cite{Mamo:2022eui}. The result of $0.68$ fm is  obtained from Ref. \cite{Guo:2021ibg}. The result of $0.67\pm0.11$ fm and $0.86\pm 0.08$ fm are extraction from charmoniums photoproduction \cite{Wang:2022vhr} and $\rho$ photoproduction \cite{Wang:2022zwz}, respectively.}  \label{fig:8}
	\end{center}	
\end{figure}

As a comparison, we directly extract the proton mass radius from the CLAS \cite{Dey:2014tfa} and LEPS \cite{LEPS:2005hax} experimental data as shown in Tab. \ref{tab:table3}. One noticed that the extracted proton mass radius values showed irregular fluctuations with increasing energy. Especially at the first energy point $W=1.98$ GeV, the extracted mass radius value has a large error bar, reflecting the large error of the experimental data at this point. The fitting results and the differential cross sections of $\phi$ photoproduction are presented in Fig. \ref{fig:9}, from which it can be seen that the data points at $W=1.98$ GeV are discretely distributed with large uncertainties. So the mass radius obtained at $W=1.98$ GeV is indeterminate. The phenomenon might be related to the difficulty of the experiment in which the energy closer to the threshold is more complicated, while $W=1.98$ GeV is very near to the threshold. Therefore, the average mass radius is $\sqrt{\left\langle r^{2}\right\rangle _\mathrm{m} }=0.80\pm0.05$ fm with $W\in[2.02,2.29]$ GeV, which is close to the result of the two gluon exchange model. Taking into account the error bar, we summarize the mass radius of the proton from this work and other groups \cite{Kharzeev:2021qkd,Wang:2022vhr,Pefkou:2021fni,Mamo:2022eui,Guo:2021ibg,Wang:2021dis,Wang:2022zwz} generally exist in $r_{m}\in[0.52,0.94]$ fm as shown in Fig. \ref{fig:8}.

It should be noticed that the closed interval does not consider the mass radius from the pomeron model.
Because the result of the pomeron model is more than $1.00$ fm and significantly larger than that of the two gluon exchange model and other groups. The primary explanation for the consequence is that the slope of the curve of the differential cross sections obtained via the pomeron model in Fig. \ref{fig:4} is massive, which leads to a larger mass radius.
So the result obtained by the pomeron model is not considered in the subsequent discussion of this paper. However, the pomeron model is still powerful for predicting the cross section of vector mesons photoproduction. In future studies, the pomeron model will be upgraded and modified to better simulate the process of vector mesons photoproduction.
\begin{table}\small
\caption{\label{tab:table3} Mass radius is derived by the differentiation experiment cross section from CLAS \cite{Dey:2014tfa} and LEPS \cite{LEPS:2005hax} collaboration, the average is $0.80\pm0.05$ fm with $W\in[2.02,2.29]$ GeV.}
	\begin{ruledtabular}
		\begin{tabular}{ccccc}
			$W$ GeV  &$1.98$ &$2.02$&$2.07$&$2.12$ \\
			\hline
			$\sqrt{\left\langle r^{2}\right\rangle _\mathrm{m}}$ fm&$0.30 \pm 0.36$&$0.80 \pm 0.07$&$0.85 \pm 0.04$&$0.95\pm0.04$\\
			\hline
			 $W$ GeV&$2.16$&$2.20$&$2.25$&$2.29$\\
			 \hline
			$\sqrt{\left\langle r^{2}\right\rangle _\mathrm{m}}$ fm&$0.80 \pm 0.05$&$0.79 \pm 0.04$&$0.68 \pm 0.05$&$0.70\pm0.04$\\
		\end{tabular}
	\end{ruledtabular}
\end{table}


\begin{figure}[htbp]
	\begin{center}
		\includegraphics[scale=0.4]{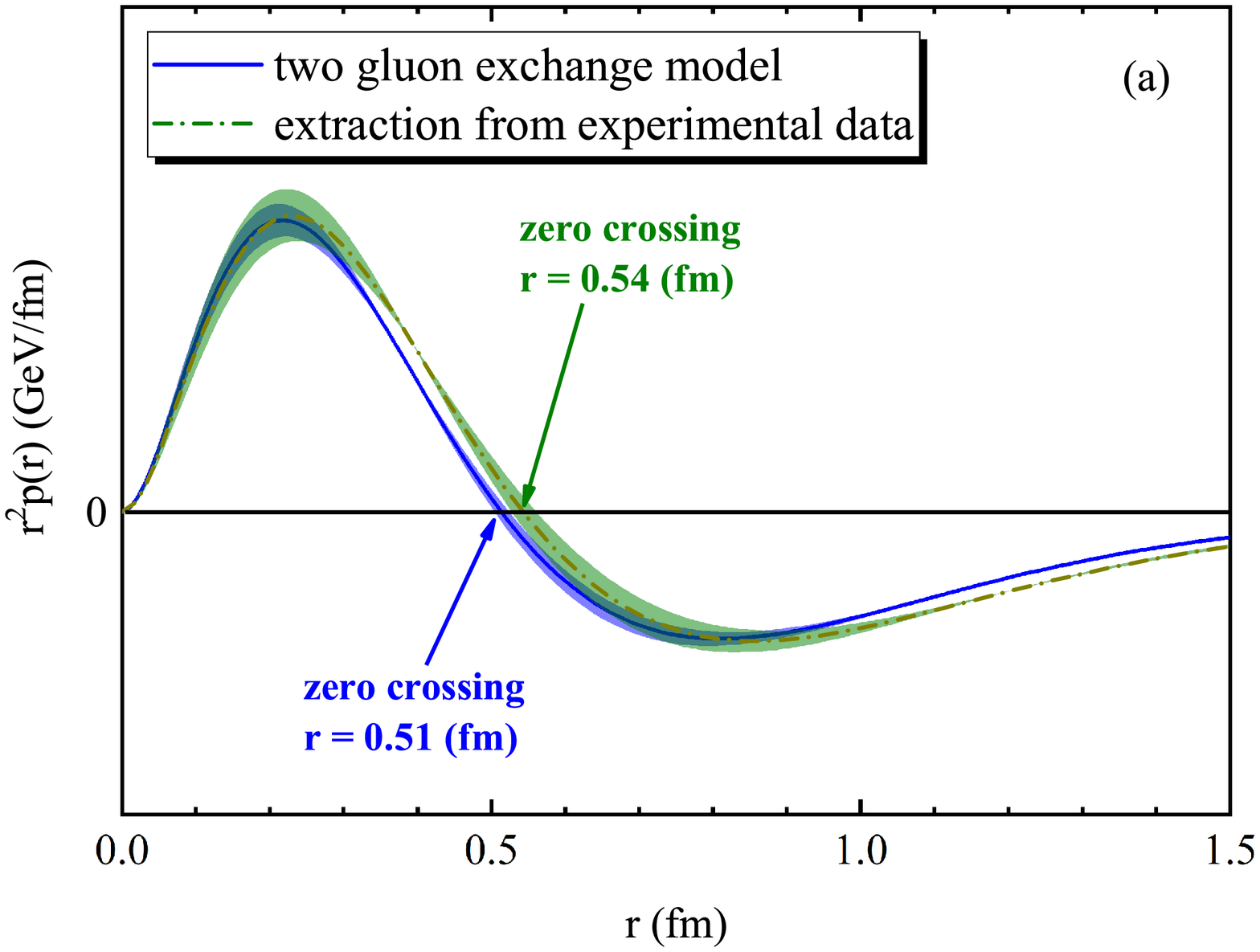}
		\\
		\includegraphics[scale=0.4]{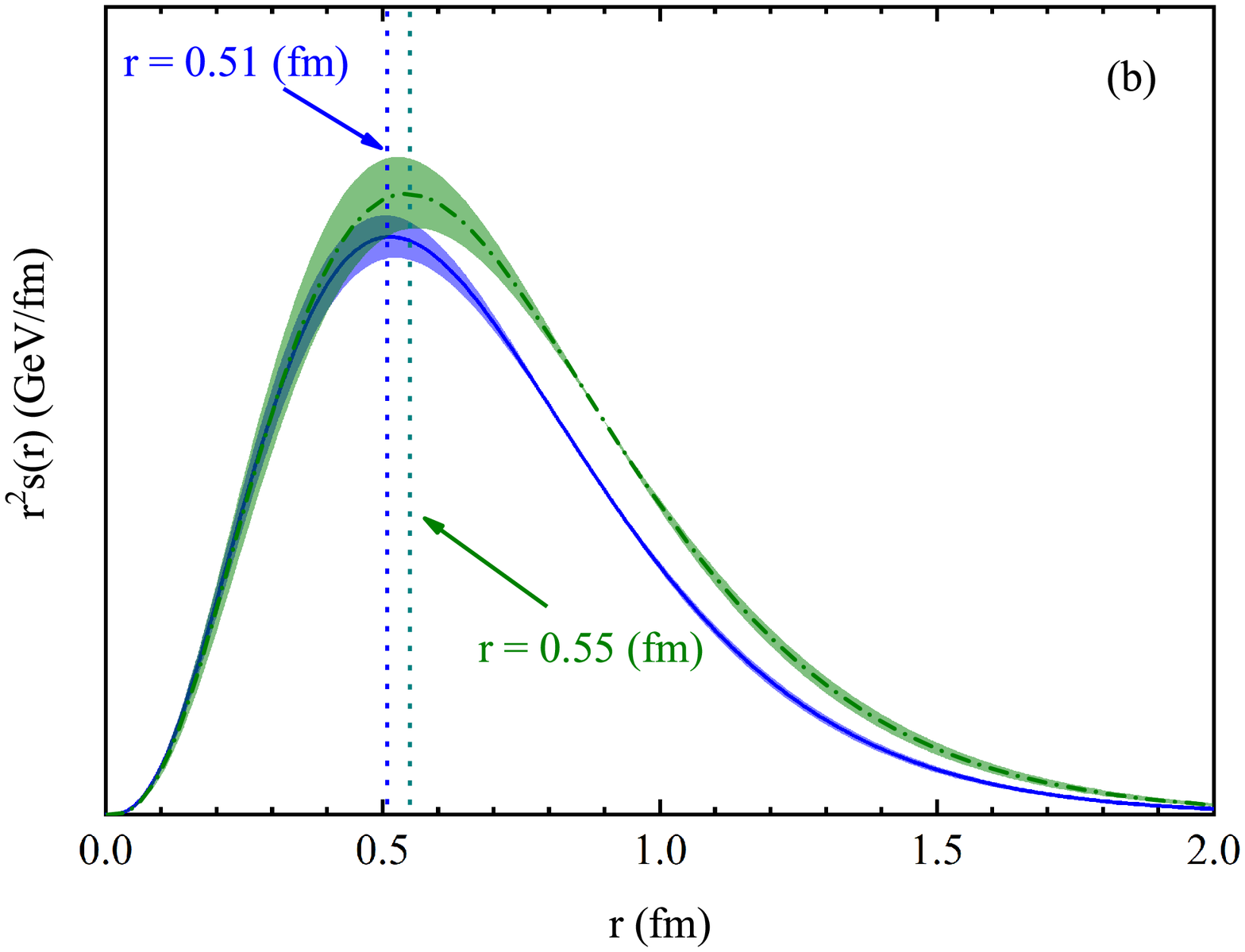}
		\caption{The internal pressure (a) and shear force (b) distribution of the proton. Here, the blue solid line is the result based on the two gluon exchange model, and the green dash-dotted line represents the result extracted from CLAS \cite{Dey:2014tfa} and LEPS \cite{LEPS:2005hax} experimental data. Here, taking the error of $m_{D}$ as the error band.}   \label{fig:10}
	\end{center}	
\end{figure}

The differential cross section of vector meson photoproduction is connected with the D-term \cite{Hatta:2021can},  which is $d \sigma / d t \propto D^{2}(t)$. So mechanical properties of the proton can be analyzed with the same data.
Then, Eq. (\ref{eq:23}) and Eq. (\ref{eq:24}) are used to calculate the pressure and shear force distributions, respectively. $D(0)$ and $m_{D}$ in $D(t)$ are free parameters that affect the size and shape of the two distributions, respectively. The shape $m_{D}$ primarily controls the distribution structure of pressure and shear force, while the size $D(0)$ only affects the distribution intensity. Therefore, exclusively $m_{D}^{2}$ is involved in the subsequent discussion.
$m_{D}^{2}$ extracted by the two gluon exchange model and from CLAS \cite{Dey:2014tfa} and LEPS \cite{LEPS:2005hax} experiment data are calculated with the analogous way of the mass radius and the result of relevant models as shown in Tab. \ref{tab:table2}. The mechanical properties based on the two gluon exchange model and the experimental data are shown in Fig. \ref{fig:10}.
\begin{figure}[htbp]
	\begin{center}
		\includegraphics[scale=0.4]{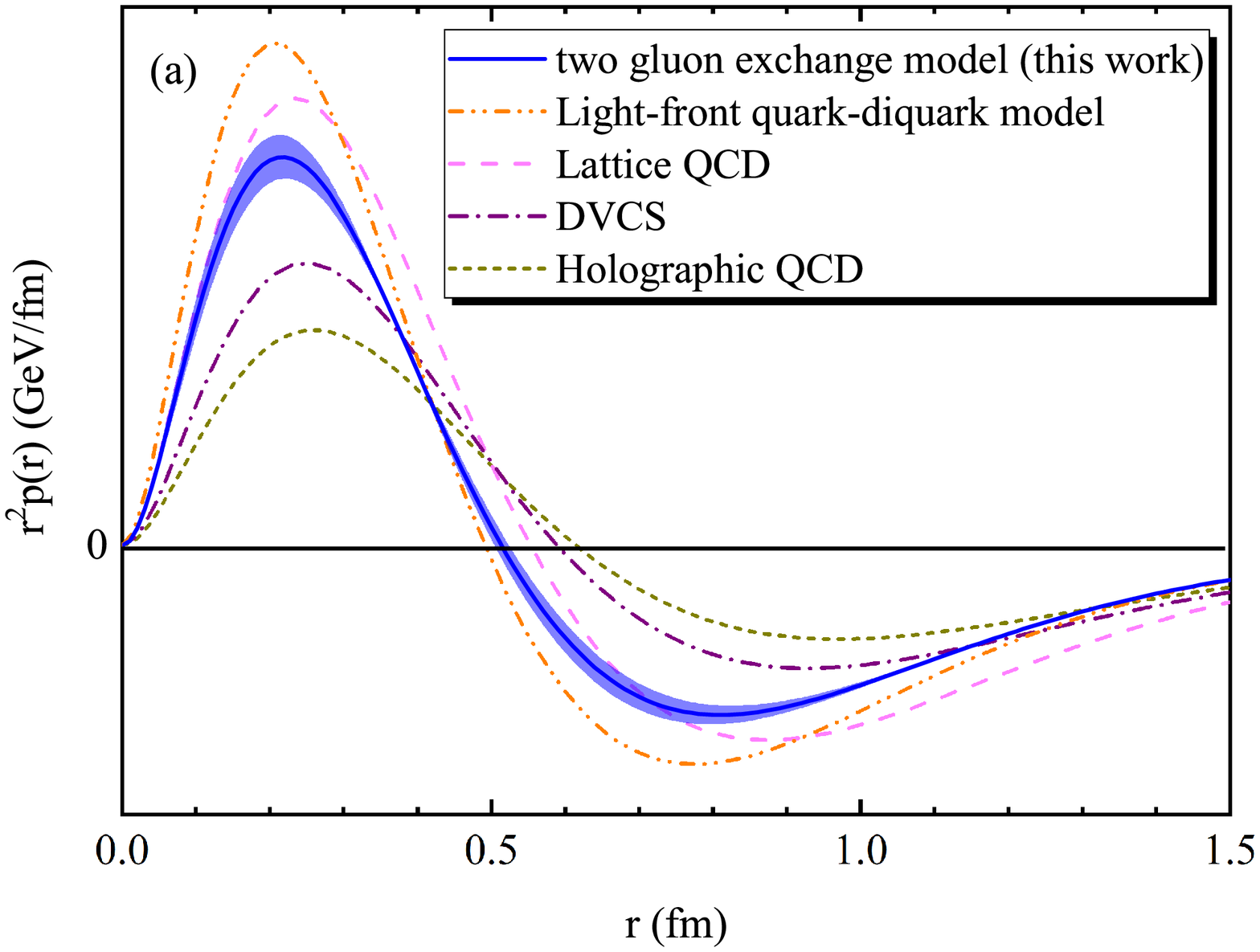}
		\\
		\includegraphics[scale=0.4]{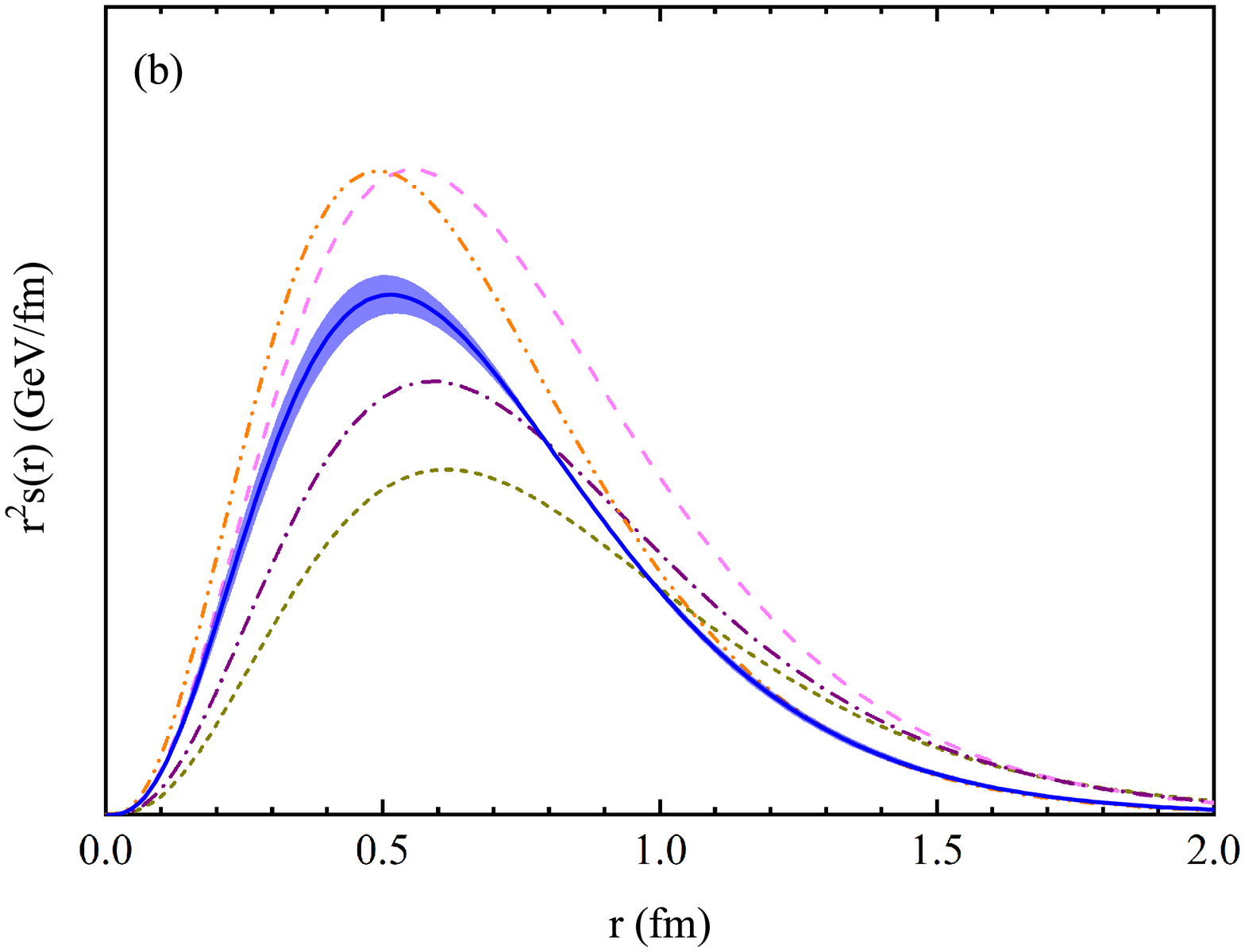}
		\caption{The internal pressure (a) and shear force (b) distribution of the proton from two gluon exchange model (blue solid line ) and other groups \cite{Pefkou:2021fni,Mamo:2022eui,Burkert:2018bqq,Burkert:2021ith,Chakrabarti:2021mfd}. The pale magenta dash line is the result of LQCD from Ref. \cite{Pefkou:2021fni}. The dark yellow dashed line and purple dash-dotted line are the result of Holographic QCD \cite{Mamo:2022eui} and DVCS \cite{Burkert:2018bqq,Burkert:2021ith} , respectively. The orange double dot dash line is the result of the light front quark-diquark model \cite{Chakrabarti:2021mfd}.}   \label{fig:11}
	\end{center}	
\end{figure}

\begin{table}\small
	\caption{\label{tab:table2} The average $ \sqrt{\left\langle m_{D}^{2}\right\rangle}$ from this work and other groups \cite{Pefkou:2021fni,Mamo:2022eui,Burkert:2021ith,Chakrabarti:2021mfd}.}
	\begin{ruledtabular}
		\begin{tabular}{cc}
			Method  &  $ \sqrt{\left\langle m_{D}^{2}\right\rangle}$\\
			\hline
			Two gluon exchange model (this work) & $1.15 \pm 0.05$ \\
			\hline
			Extraction from CLAS and LEPS data (this work)&$1.09\pm0.07$\\
			\hline
			Lattice QCD \cite{Pefkou:2021fni}&$1.07$\\
			\hline
			Holographic QCD \cite{Mamo:2022eui}&$0.963$ \\
			\hline
			DVCS \cite{Burkert:2021ith}&$1.01 \pm 0.13$\\
			\hline
			Light-front quark-diquark model \cite{Chakrabarti:2021mfd} & $1.37$
		\end{tabular}
	\end{ruledtabular}
\end{table}

Initiating from the center of the proton $r=0$ fm, the radial distribution of pressure $r^{2}p(r)$ is divided into positive and negative parts by zero-crossing, dominated by repulsive force and binding force.
It can be observed from Fig. \ref{fig:10} that the zero-crossing based on the experiment data has a bit larger than that on the two gluon exchange model, and the positive and negative peaks are also more sizeable. What is striking is that the zero-crossing are significantly adjacent, which is similar to the result of mass radius.
The peak of the radial distribution of shear force $r^{2}s(r)$ has the same trend as that of pressure. The result developed by other groups \cite{Pefkou:2021fni,Mamo:2022eui,Burkert:2018bqq,Burkert:2021ith,Chakrabarti:2021mfd} are compared with that of the two gluon exchange model, as shown in Fig. \ref{fig:11}. It is found that the distribution results based on the two gluon exchange model are roughly in line with those of the other groups, but there are some differences. In fact, the current research on the internal mechanical properties of the proton is still in the early stage, including the different contributions of quarks and gluons to the distribution of mechanical properties given by different models or experiments. Therefore, we need to give the results of the corresponding proton mechanics properties from different models or experiments, so as to facilitate the in-depth study of this problem in the future.

\begin{figure}[htbp]
	\begin{center}
		\includegraphics[scale=0.4]{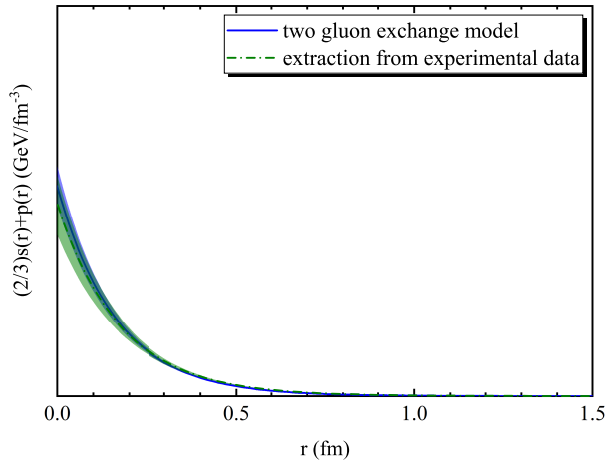}
		\caption{Schematic diagram of $(2/3)s(r)+p(r)$ as a function of $r$ for a mechanical system satisfies the stability conditions. Here, the notations are the same as in Fig. \ref{fig:10}.}   \label{fig:13}
	\end{center}	
\end{figure}

In addition, it can be seen from Tab. \ref{tab:table2} that the zero-crossing point decreases with the increase of $ \sqrt{\left\langle m_{D}^{2}\right\rangle}$.
The stability condition of the mechanical system argued in Ref. \cite{Polyakov:2018zvc} is $(2/3)s(r)+p(r)>0$. That is, the corresponding force must be outward. As shown in Fig. \ref{fig:13}, the mechanical system based on the two gluon exchange model and the the result extracted from the $ \phi$ photoproduction data satisfies the inequality. However, although the mechanical properties satisfy the mechanical stability condition of the system, and the radial distribution agrees with other models, there still exist uncertain factors in the proton mechanical properties, among which $D(0)$ is one.

\begin{figure}[htbp]
  \centering
  \subfigure[]{\includegraphics[scale=0.75]{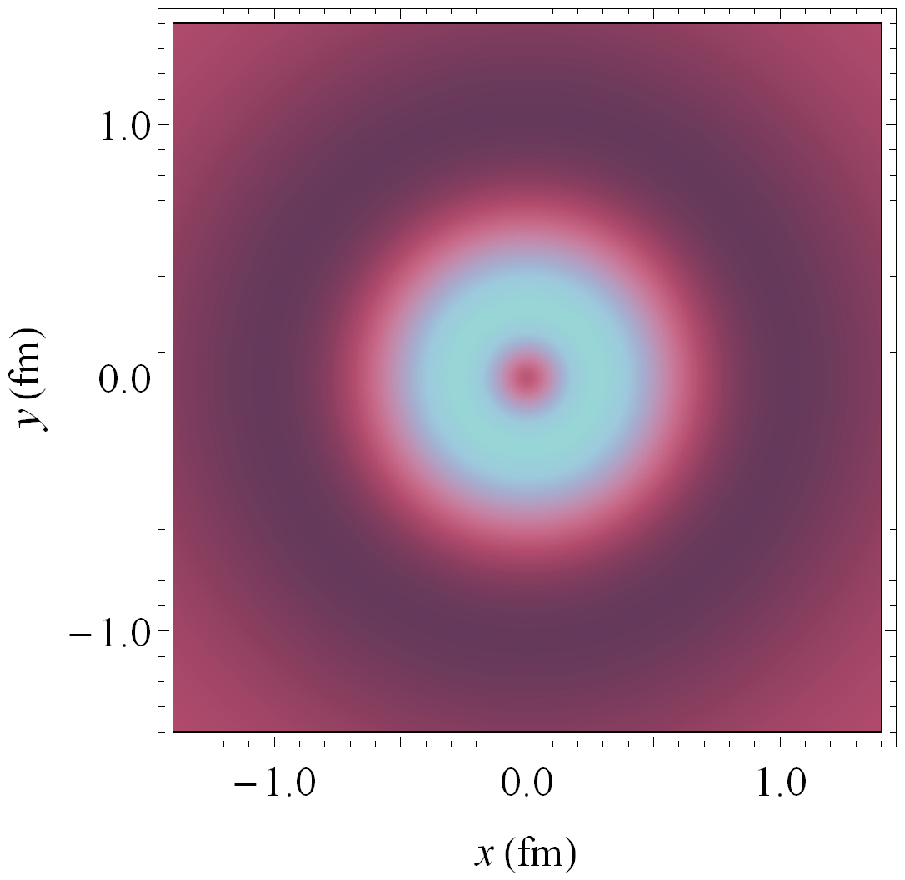}
  \includegraphics[scale=0.73]{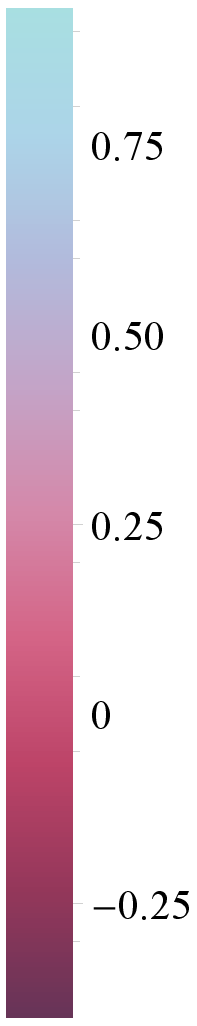}
  }
  \\
  \subfigure[]{
  \includegraphics[scale=0.70]{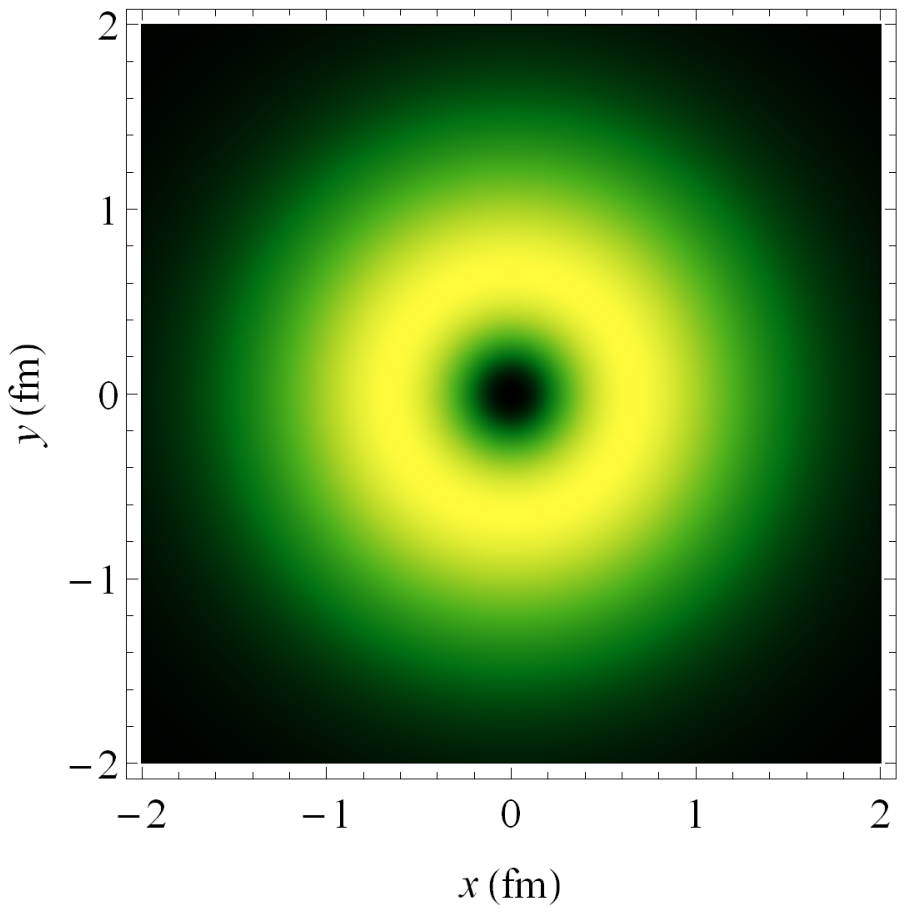}
  \includegraphics[scale=0.71]{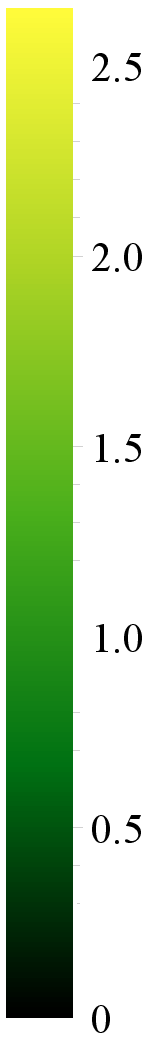}
  }
  \caption{2D display of the pressure $r^{2}p(r)$ (a) and shear force $r^{2}s(r)$ (b) distribution in proton from the proton center based on the two gluon exchange model.}
  \label{fig:12}
\end{figure}

2D displays of pressure and shear force are established based on the two gluon exchange model in Fig. \ref{fig:12}, which describes the pressure and shear force distribution from the proton center. The colorbar on the Fig.\ref{fig:12} (a) describes the areas above $0$ where repulsive forces dominate and below $0$ where binding forces dominate. With $0$ as the node, the intensity increases first and then decreases as the distance from the center of mass increases. The color bar on Fig.\ref{fig:12} (b) describes the variation of the strength of the shear force distribution as the distance from the center of mass increases.

\section{Summary} \label{sec4}
This work reproduces the differential and total cross sections of $\gamma p \rightarrow \phi p$ at the near-threshold based on the two gluon exchange model with the gluon distribution function and pomeron model developed by larget \cite{Laget:1994ba}. The parameters $A_{0}, A_{1}, A_{2}$ as well as $b_{0}$ are obtained by fitting differential cross sections from   CLAS \cite{Dey:2014tfa} and LEPS \cite{LEPS:2005hax} and total cross section from CLAS \cite{Barber:1981fj,Strakovsky:2020uqs}, SLAC \cite{Egloff:1979mg}, FERMI \cite{Busenitz:1989gq,Owens:2012bv}, LAMP2 \cite{Ballam:1972eq}. The average mass radius from the two gluon exchange and pomeron models are $0.78 \pm 0.06$ fm and $1.02 \pm 0.01$ fm, respectively. In addition, the mass radius is analyzed with the CLAS \cite{Dey:2014tfa} and LEPS \cite{LEPS:2005hax} experimental data to be $0.80\pm0.05$ fm which is close to the result of the two gluon exchange model. Based on the mass radius obtained by this work, and ignoring the conclusions of the pomeron model, we concluded that the mass radius of the proton given by the $\phi$ meson photoproduction process should be attributed in $r_{m}\in[0.78,0.80]$ fm. At present, the proton mass radius value extracted by the cross section of heavy charmoniums photoproduction is roughly between 0.55 and 0.68 fm. In our recent work \cite{Wang:2022zwz}, the proton mass radius was $0.86\pm 0.08   \text{ fm }$ extracted from the lighter $\rho$ meson photoproduction data. Interestingly, the mass of $\phi$ is between the charmoniums and the $\rho$ meson, and the corresponding proton mass radius value is also roughly in the middle. In fact, if we assume that for lighter vector mesons, the contribution from the gluon operator in the production amplitude is gradually diminishing, while the contribution from the light quark operator dominates, resulting in an increasing proton mass radius extracted and is close to the value of the proton charge radius. If this assumption holds, it means that the proton mass radius extracted by the vector meson photoproduction have some rationality. However, we must note that the proton mass radius values obtained so far are with error bars, and if we take the error into account, it is difficult to give a clear size relationship. Therefore, high-precision experimental measurement data and more complete theoretical models are very much needed.

Afterwards, taking a similar approach, we systematically investigated the mechanical properties of the proton and compared the obtained results with those of other groups. The results show that the proton internal pressure and shear force distributions given by us are close to the results of light-front quark-diquark model \cite{Chakrabarti:2021mfd} and LQCD \cite{Pefkou:2021fni}, but there is a little difference with DVCS \cite{Burkert:2018bqq,Burkert:2021ith} and holographic QCD \cite{Mamo:2022eui}. In fact, the results of proton mechanics properties given by different models or experiments are not the same. Therefore, our study may provide necessary reference or supplementary information for subsequent in-depth research on the internal properties of protons.


The Electron-Ion Collider (EIC) is a powerful facility for studying nucleons' structure and strong interactions in non-perturbation energy regions. At present, EIC facilities \cite{Accardi:2012qut,Anderle:2021wcy} in China and the United States take the study of nuclear structure and properties as an important scientific goal, and our theoretical results on the photoproduction of $\phi$ meson may provide an important reference and basis for further accurate measurements of $\phi$ meson electroproduction at EIC facilities \cite{Accardi:2012qut,Anderle:2021wcy}.

\begin{acknowledgments}
X.-Y. Wang would like to acknowledge Prof. Dmitri E. Kharzeev for useful discussions about proton mass radius. This project is supported by the National Natural Science Foundation of China under Grant Nos. 12065014 and 12047501, and by the West Light Foundation of The Chinese Academy of Sciences, Grant No. 21JR7RA201.
\end{acknowledgments}

\end{document}